\documentclass[twocolumn,showpacs,preprintnumbers,amsmath,amssymb]{revtex4}

\usepackage{graphicx}
\usepackage{dcolumn}
\usepackage{bm}
\usepackage[latin1]{inputenc}

\begin{document}

\preprint{APS/123-QED}

\title{Critical behavior of ferromagnetic pure and random diluted nanoparticles with competing interactions: variational and Monte Carlo approaches}

\author{E. A. Vel\'asquez}
 \email{eavelas@gmail.com}
 \affiliation{Grupo de Investigaci\'on en Modelamiento y Simulaci\'on Computacional, Facultad de Ingenier\'{i}as, Universidad de San Buenaventura Sede Medell\'{i}n, A.A. 5222 Medell\'{i}n Colombia.\\Grupo de Magnetismo y Simulaci\'on G+ and Grupo de Instrumentaci\'on Cient\'ifica y Microelectr\'onica, Instituto de F\'isica, Universidad de Antioquia. A.A. 1226, Medell\'in, Colombia\\
}%

\author{J. Mazo-Zuluaga}
 \email{jomazo@fisica.udea.edu.co}
\author{J. Restrepo}
 \email{jrestre@fisica.udea.edu.co}
\affiliation{Grupo de Magnetismo y Simulaci\'on G+ and Grupo de Instrumentaci\'on Cient\'ifica y Microelectr\'onica, Instituto de F\'isica, Universidad de Antioquia. A.A. 1226, Medell\'in, Colombia\\
}%
\author{\`{O}scar Iglesias}
 \email{oscar@ffn.ub.es}
 \affiliation{Departament de F\'isica Fonamental and Institut de Nanoci\`encia i Nanotecnologia, Universitat de Barcelona, Diagonal 647, 08028 Barcelona, Spain}

\date{\today}

\begin{abstract}
The magnetic properties and critical behavior of both ferromagnetic pure and metallic nanoparticles having concurrently atomic disorder, dilution and competing interactions, are studied in the framework of an Ising model. We have used both the free energy variational principle based on the Bogoliubov inequality and Monte Carlo simulation. As a case of study for random diluted nanoparticles we have considered the Fe$_{0.5}$Mn$_{0.1}$Al$_{0.4}$ alloy characterized for exhibiting, under bulk conditions, low temperature reentrant spin glass (RSG) behavior and for which experimental and simulation results are available. Our results allow concluding that the variational model is successful in reproducing features of the particle size dependence of the Curie temperature for both pure and random diluted particles. In this last case, low temperature magnetization reduction was consistent with the same type of RSG behavior observed in bulk in accordance with the Almeida-Thouless line at low fields. A linear dependence of the freezing temperature with the reciprocal of the particle diameter was also obtained. Computation of the correlation length critical exponent for random diluted nanoparticles yielded the values $\nu$=0.926$\pm$0.004 via Bogoliubov and $\nu$=0.71$\pm$0.04 via Monte Carlo. Differences are attributed to the so-called pairs-approximation in the variational model. From both approaches, differences in the $\nu$ exponent of Fe$_{0.5}$Mn$_{0.1}$Al$_{0.4}$ nanoparticles with respect to that of the pure Ising model agree with Harris and Fisher arguments.
\end{abstract}

\maketitle

\section{Introduction}
The magnetic properties of bulk metallic systems having concurrently atomic disorder, dilution, competing interactions and characterized for exhibiting spin glass (SG) behavior have been widely studied from different points of view: experiment, theory and numerical simulation \cite{binder-young,parisi,young}. Among the systems with such characteristics, we can mention for instance CuMn \cite{mydosh}, FeAu \cite{khmelevskyi}, FeAl \cite{raymond,aguirre,zamora,lue}, FeNiMn \cite{pappas,penha} and FeMnAl \cite{penha,restrepo,restre-greneche,restre-perez,osorio} alloys. These alloys are interesting due to the richness of magnetic phases that can be found such as ferromagnetic, antiferromagnetic, superparamagnetic, cluster glass, SG and RSG depending on stoichiometry, microstructure, degree of dilution, atomic disorder, magnetic field and temperature. Typical Ising spin glass systems like those based on FeMnTiO$_3$ \cite{torikai,jonsson}, Fe(Cu,Al)Dy \cite{imai}, LiHoYF \cite{ka-ming}, FeAl \cite{raymond,aguirre,lue,shukla} and FeMnAl \cite{penha,restrepo,osorio} are good candidates to study SG and RSG related properties through Ising-based theoretical models where good agreement with experimental results has been achieved. In particular, pure SG and RSG behaviors, in ternary FeMnAl alloys, arise from several ingredients including random atomic distribution of the alloy constituent elements in the crystalline structure, dilution provided by Al atoms giving rise to bond randomness and, finally, competition among the different exchange integrals involved. On this last respect, competition is given, essentially, by the difference in sign and magnitude of the $J_\text{{Fe-Fe}}$, $J_\text{{Fe-Mn}}$ and $J_{\text{Mn-Mn}}$ exchange integrals. For high enough iron contents, a RSG behavior within the ferromagnetic phase governed by the Fe matrix can arise \cite{restre-greneche}.\\
Up to date, works reported on these kind of alloys deal with magnetic properties under bulk conditions. However, to our knowledge, and despite of all the literature related with the so-called surface spin glass-like behavior in nanoparticles, no studies on metallic nanoparticles having concurrently bond competition, magnetic dilution and atomic disorder within the entire volume of the nanoparticles have been reported. This fact has led us to consider the interplay between these effects and those arising from finite size when considering nanoparticles having such ingredients. 
The purpose of this article is to characterize from the magnetic standpoint, how the SG behaviors, found in systems like FeMnAl alloys under bulk conditions, become revealed in nanoparticles where the surface to volume ratio becomes increasingly important. Both the free energy variational method based on the Bogoliubov inequality and a Metropolis Monte Carlo simulation in the framework of a nearest neighbor  Ising model were considered. The former, as energy minimization tool, has been already successfully employed in describing the magnetic properties of this kind of systems where theoretical magnetic phase diagrams are in good agreement with the experimental ones \cite{penha,restrepo,osorio}. The layout of the paper is as follows. In Sec. II we describe the theoretical model and we emphasize the importance of a relationship for the average nearest neighbor coordination number as function of the particle size. In Sec. III we present our numerical results. This section provides finite-size scaling analysis of both pure ferromagnetic nanoparticles, with an application to Ni nanostructures, and  Fe$_{0.5}$Mn$_{0.1}$Al$_{0.4}$ nanoparticles from both approaches. Conclusions are finally presented in Sec. IV.

\section{Theoretical model}
Several features lead us to consider an Ising model: i) magnetic frustration can be better resolved with an Ising model than, for instance, with continuous spin models \cite{binder-young}, ii) it is in agreement with the framework we are interested in, which consists of iron-based nanoparticles with a very high effective magnetocrystalline anisotropy and where, despite of having cubic structure, a single easy axis can be experimentally induced \cite{getzlaff,arantxa,kleibert}, iii) it has been already used in similar systems and quite good agreement with experimental data (magnetometric measurements and hyperfine fields from M\"ossbauer spectroscopy) has been achieved \cite{penha,aguirre,restrepo,osorio}, and finally iv) it allows to keep computational requirements under reasonable limits. Thus, our model is based on the following $N$ spins Ising Hamiltonian:
\begin{equation}
\label{1}
{\cal H}= -\sum_{\left\langle i,j\right\rangle} J_{ij} \sigma _{i} \sigma_{j} - h\sum^{N}_{i=1} \sigma _{i}.
\end{equation}
The first sum runs over nearest neighbors $\left\langle i,j\right\rangle$, and $\sigma_i$ takes on the values $\pm 1$ or $0$ depending on whether the $i^{th}$ site is occupied by a magnetic atom (Fe,Mn) or an aluminum one, respectively. The exchange integral $J_{ij}$ obeys the following probability distribution function accounting for disorder and the different couplings involved \cite{restrepo}:
\begin{eqnarray}
 \label{2}
P(J_{ij}) &=& p^{2} \delta (J_{ij} - J_\text{{Fe-Fe}}) + 2px\delta (J_{ij} - J_\text{{Fe-Mn}})\\ \nonumber &+& x^{2}\delta (J_{ij} - J_\text{{Mn-Mn}}) + (q^{2}+2xq+2pq)\delta (J_{ij}),
\end{eqnarray}
where $p$, $x$ and $q$, with $p+x+q=1$, are the fractional concentrations of Fe, Mn and Al atoms, respectively. The terms $p^2$, $2px$ and $x^2$ are the probabilities of having nearest neighbors Fe-Fe, Fe-Mn and Mn-Mn bonds, respectively, and interacting through the corresponding exchange integrals $J_\text{{Fe-Fe}}$, $J_\text{{Fe-Mn}}$ and $J_\text{{Mn-Mn}}$. Here, $J_\text{{Fe-Fe}}$, hereafter simply $J$, was set to 12.846 meV only for pure iron nanoparticles with body-centered cubic (bcc) structure, and was set to 16.872 meV for Fe$_{0.5}$Mn$_{0.1}$Al$_{0.4}$ nanoparticles having the same structure. These values reproduce the Curie critical temperatures of the corresponding systems under bulk conditions and the difference among them is attributed to the presence of both Mn and Al atoms and to the fact that Fe and Fe$_{0.5}$Mn$_{0.1}$Al$_{0.4}$ have different lattice parameters \cite{restre-perez}. Additionally, for the alloy, the remaining exchange integrals were set to $J_\text{{Fe-Mn}}$=$-\alpha J$ and $J_\text{{Mn-Mn}}$=$- \lambda J$ with $\alpha$=0.005  and $\lambda$ =0.03, \cite{restrepo} and they correspond to the so-called competitive parameters. In this work, such values are kept fixed regardless the size of the nanoparticles to be considered. The last coefficient $q^2+2xq+2pq$ stands for diluted bonds, with $J_{ij}$=0, corresponding to nearest neighbors Al-Al, Al-Mn and Al-Fe pairs. Finally, the second term in Eq.(\ref{1}) is the Zeeman contribution dealing with the coupling of the spins with a uniform external applied magnetic field $h$.\\Following the ideas proposed by Ferreira \textit{et al.} \cite{ferreira}, in the pairs-approximation, the system is considered as formed by $n_1$ single spins (S) and $n_2$ linked pairs (P) of spins with a total number of spins $N=n_1+2n_2$. Additionally, it is assumed that the magnetization can be obtained either from single spins or spins belonging to a pair. Thus, the trial Hamiltonian can be written as:
\begin{equation}
 \label{3}
{\cal H}_{0}= -\gamma_{s}\sum_{i\in\text{S}}\sigma_{i} - \sum_{j, k \in\text{P}} \left[J_{ik}\sigma_{j}\sigma_{k} + \gamma_{p}\left(\sigma_{j}+\sigma_{k}\right)\right],
\end{equation}
where $\gamma_s$ and $\gamma_p$ are variational parameters which can be interpreted as molecular fields to be determined from energy minimization conditions. Here, the first sum runs over single spins and the second sum runs over pairs. Both Hamiltonians, Eqs.(\ref{1}) and (\ref{3}), can be related through the variational approach based on the Bogoliubov inequality:
\begin{equation}
 \label{4}
\left[F\right] \leq \left[F_{0}\right] +  \left[\left\langle  {\cal H} - {\cal H}_{0} \right\rangle_{0}\right]\equiv \left[\Phi\right],
\end{equation}
where $F$ is the Helmholtz free energy defined by ${\cal H}$, $F_0$ is the free energy defined by ${\cal H}_{0}$; $\left\langle ... \right\rangle_{0}$ refers to the thermal average in the ensemble defined by ${\cal H}_{0}$, whereas $\left[ ... \right]$ represents a configurational average in which atomic disorder is considered. According to the way as the system has been figured out we have:
\begin{equation}
 \label{5}
F_{0} = - k_B T ln Z_{0}=- k_B T ln (Z_{s}^{N-2n_2} Z_{p}^{n_2}),
\end{equation}
where
\begin{equation}
 \label{6}
Z_{s} = 2 cosh (\beta \gamma_{s}),
\end{equation}
\begin{equation}
 \label{7}
Z_{p} = 2 e^{\beta J_{ij}} cosh (2 \beta \gamma_{p}) + 2 e^{-\beta J_{ij}},
\end{equation}
are the trial partition functions for single and pair spins, respectively. The configurational average  of any observable $A$ is obtained from:
\begin{equation}
 \label{8}
\left[A\right] = \int_{\{J_{ij}\}}AP(J_{ij})dJ_{ij}.
\end{equation}
Calculation of the quantities $\left[F_{0}\right]$, $\left[\left\langle {\cal H}-{\cal H}_{0}\right\rangle_{0} \right]$ and $\left[\Phi\right] $, following the same procedure as it has been described elsewhere \cite{restrepo,ferreira}, leads to the following expression for energy minimization:
\begin{eqnarray} \nonumber
  \frac{\partial \left[\Phi\right]}{\partial m} = &-& 2\left(n'-n_{2}\right)(p^{2}-2px\alpha  - \lambda x^{2})J m\\  
  &-& Nh + \left( N-2n_{2}\right)\gamma_{s}+ 2n_{2}\gamma_{p}= 0,\label{9}
  \end{eqnarray} 
where $n'$ is the number of nearest neighbors, which depends on the crystalline structure, the type of boundary conditions and the system size. As concerns to nanoparticles we consider free boundary conditions.
Since $\left[\Phi\right]$ diminishes as $n_2$ increases, we take $n_2$ as large as physically possible, i.e. $n'=n_2$. Thus, the number of linked pairs is maximized. Hence we obtain the following relationship between the variational parameters or molecular fields $\gamma_s$ and $\gamma_p$:
\begin{equation} \label{10}
\gamma_{s}= \frac{(2 \gamma_{p}n'/N)- h}{(2n'/N)-1}.
\end{equation}
Our system is a spherical nanoparticle composed by $N$ atoms arranged in a bcc structure with core coordination number $z$=8 and interatomic spacing $a$, which in the case of pure iron is around 2.86\AA\  whereas for Fe$_{0.5}$Mn$_{0.1}$Al$_{0.4}$ is around 2.96\AA \cite{restre-perez}. On the basis of such representation and in order to get an expression for the maximum number of pairs divided by the total number of atoms, i.e. $n'/N$, or, analogously an effective coordination number $z_\text{eff}=2n'/N$, we have simulated particles with bcc structure and different diameters $D$, in units of the lattice parameter $a$, and have counted the number of nearest neighbors pairs. Figure 1 shows the size dependence of $z_\text{eff}$ from which the following relationship is fulfilled for $D\geq3.4$: 
\begin{equation}
\label{11}
z_\text{eff}=\frac{2n'}{N}= z-\frac{b}{D},
\end{equation}
where $z$=8 and the best fit yields $b=10.13\pm0.04$ for a bcc lattice. In the case of a face-centered cubic (fcc) lattice, the following values must be used, $z$=12 and $b=12.65\pm0.06$. In principle, $z_\text{eff}$  can be interpreted as an effective coordination number for nanoparticles of diameter $D$ whose core coordination number is $z$.
\begin{figure}[h]
 \centering
 \includegraphics[scale=0.85]{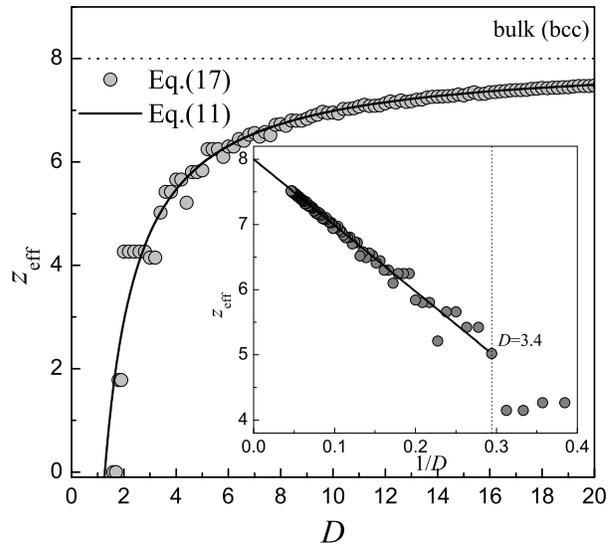}
 \caption{\label{fig01} Size dependence of the effective (Eq.(\ref{11})) and the average (Eq.(\ref{17})) coordination number for nanoparticles having bcc structure. Particle diameter $D$ is given in units of the lattice parameter $a$.}
\end{figure}
Thus, the relationship between the molecular fields $\gamma_s$ and $\gamma_p$ can be rewritten as:
\begin{equation} \label{12}
\gamma_s=\frac{\gamma_p z_\text{eff}-h}{z_\text{eff}-1}
\end{equation}
Magnetization can be computed either from single spins or from spins linked to a pair, and it must be the same:
\begin{equation} \label{13}
m=\frac{1}{\beta}\frac{\partial ln Z_{s}}{\partial \gamma_{s}}=\frac{1}{2\beta}\frac{\partial ln Z_{p}}{\partial \gamma_{p}}.
\end{equation}
After calculating the derivatives and performing the configurational averages using the bonds probability distribution function given by Eq.(\ref{2}), we obtain the following transcendental equation for the magnetization:
\begin{eqnarray} \nonumber
m &=&tanh (\beta \gamma_{s}) \\ \nonumber
&=& sinh(2\beta\gamma_{p}) \left[ \frac{p^{2}}{cosh(2\beta\gamma_{p})+ e^{-2\beta J}}\right. \\ \nonumber
&&+   \frac{2px}{cosh(2\beta\gamma_{p})+ e^{2\alpha\beta J}} \\ 
&&+ \left. \frac{x^{2} }{cosh(2\beta\gamma_{p})+e^{2\lambda\beta J}} + \frac{q^{2}+2pq+2qx}{cosh(2\beta\gamma_{p})+1}\right]. \label{14}
\end{eqnarray}
Roots of this equation were obtained by using the FindRoot tool of Mathematica\texttrademark. It must be stressed that such magnetization, according to the bonds distribution function in Eq.(\ref{2}), corresponds to an average magnetization per bond whereas for the pure case ($p$=1) becomes a magnetization per site. The presence of crossed terms involving the atomic concentrations of the constituent elements $p$, $x$ and $q$ in Eq.(\ref{14}) reflects the average over all possible nearest neighbors pairs as well as the random atomic distribution feature. Zero field magnetic susceptibility was obtained according to $\chi=\left(\partial m / \partial h\right)_0$, yielding:
\begin{eqnarray} \nonumber
\chi  = & & \left( \frac{\partial m}{ \partial h} \right)_0 = \left\{ (1-z_{\text{eff}}) \left(\beta sech^{2}(\beta \gamma_{s}) \right)^{-1} \right. \\ \nonumber
&+& \frac{z_{\text{eff}}}{2 \beta} \left[tanh(\beta \gamma_{s})coth(2\beta \gamma_{p})   \right. \\ \nonumber
&-&   sinh^{2}(2\beta \gamma_{p})\left( \frac{p^{2}}{(cosh(2\beta\gamma_{p})+ e^{-2\beta J})^{2}} \right.\\ \nonumber
&+&\frac{2 p x }{(cosh(2\beta\gamma_{p})+e^{2\alpha\beta J})^{2}}+ \frac{x^{2} }{(cosh(2\beta\gamma_{p})+ e^{2\lambda\beta J})^{2}} \\ 
&+& \left. \left. \left.  \frac{q^{2}+2pq+2qx}{(cosh(2\beta\gamma_{p})+1)^{2}}\right) \right]^{-1}
\right\}^{-1} \label{15}
\end{eqnarray}\\
The calculation of the Curie temperature $T_C$ from Eq. (\ref{14}), for which $m$=0, is performed by taking the limits $\gamma_p\to$0 and $\gamma_s\to$0. This yields the following expression for the magnetic phase diagram:
\begin{eqnarray} \nonumber
\frac{z_\text{eff}}{2(z_\text{eff}-1)}=\frac{p^2}{1+e^{-2\beta_CJ}}+\frac{2px}{1+e^{2\alpha\beta_CJ}}+\\
\frac{x^2}{1+e^{2\lambda\beta_CJ}}+\frac{q^2+2pq+2qx}{2}\label{16}
\end{eqnarray}
Here, $\beta_C=(k_BT_C)^{-1}$. 

\section{Results and discussion}
\subsection{Coordination number}
From the particle size dependence of $z_\text{eff}$ plotted in Fig. 1, particles with diameter $D$=10 (around 3 nm) already exhibit an effective coordination number around 7, which corresponds roughly to 88\% of that of the bulk. Above $D$=10, the effective coordination number resembles that of the system under bulk conditions. Below that value, coordination number decreases rapidly and, therefore, strong modifications on the magnetic properties are expected to occur in this range, i.e. below around 3 nm. In order to gain a deeper insight on the interpretation of $z_{\text{eff}}$, we have computed the coordination number per particle $z_i$ by counting the number of nearest neighbors surrounding the atom at the $i^{th}$ position, i.e. within the first coordination shell, and an average coordination number was computed according to:
 \begin{equation} \label{17}
\left< z\right>=\frac{1}{N}\sum_{i=1}^{N}n_iz_i,
 \end{equation}
where $n_i$ is the number of atoms having coordination number $z_i$ (see Fig. 2). Hence, by comparing the results derived from Eqs. (11) and (17) in Fig.1, we conclude that what we have called an effective coordination number can indeed be considered as an average coordination number, i.e. $z_\text{eff}=\left< z\right>$. For diameters below around 3.4$a$ the average coordination number is characterized by jumps for which the discrete character of the system becomes more evident. A typical particle with $D$=10 is illustrated in Fig. 2, where surface atoms with different coordinations are depicted with different colors. Our results are in agreement with those reported in metal Pt nanoparticles with different diameters where coordination numbers for the first through fifth coordination shells were obtained by extended X-ray absorption fine structure (EXAFS) spectroscopy \cite{frenkel}.
\begin{figure}[h]
 \centering
 \includegraphics[scale=0.4]{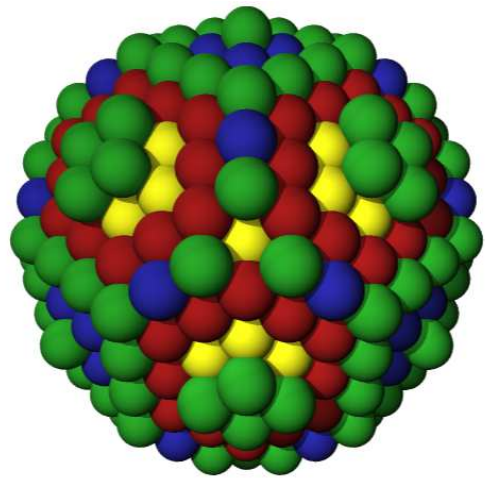}
 \caption{\label{fig02} (Color online) Particle with diameter $D$=10 having 1067 atoms, bcc structure, and an average coordination number $\left<z\right>$=6.96. Surface atoms with 4, 5, 6 and 7 nearest neighbors are colored green, blue, red, and yellow, respectively.}
\end{figure}

\subsection{Pure nanoparticles}
The particular case of pure ferromagnetic nanoparticles is easily obtained by setting $p$=1, $q$=0 and $x$=0 in Eq.(\ref{16}):
\begin{equation}\label{18}
k_BT_C(D)=\frac{2J}{ln\left[z_\text{eff}/\left(z_\text{eff}-2\right)\right]},
\end{equation}
which gives the particle size dependence of the Curie temperature for nanoparticles with core coordination $z$, diameter $D$ and a nearest neighbors exchange integral $J$. We want to stress that these results are not exclusively applicable to pure $\alpha$-Fe nanoparticles and they can be, in principle, employed for other pure ferromagnetic nanoparticles like Ni or Co. We can also relate this critical temperature with that of the system under bulk conditions \cite{restrepo} in order to obtain a reduced critical temperature:
\begin{equation}\label{19}
\frac{T_C(D)}{T_C(\infty)}=\frac{J(D)ln\left[z/(z-2)\right]}{J(\infty)ln\left[z_\text{eff}/(z_\text{eff}-2)\right]},
\end{equation}
where we have assumed that the exchange integral ($J=J(D)$) in the nanoparticle can be different from that under bulk conditions ($J(\infty)$). In a first order of approximation, and by assuming the same exchange integral value, which could be reasonable for high enough particle sizes, we have:
\begin{equation}\label{20}
\frac{T_C(D)}{T_C(\infty)}\approx\frac{z_\text{eff}}{z},
\end{equation}
if we assume that no structural transition occurs as a consequence of reducing the size. Otherwise, different core coordination numbers should be considered and the model is still applicable. Figure 3 shows the reduced critical temperature for different diameters according to our model. A comparison between Eq.(\ref{19}) for bcc ($z$=8) and fcc ($z$=12) lattice structures and the approximate expression given by Eq.(\ref{20}) is included.
\begin{figure}[h]
 \centering
 \includegraphics[scale=0.85]{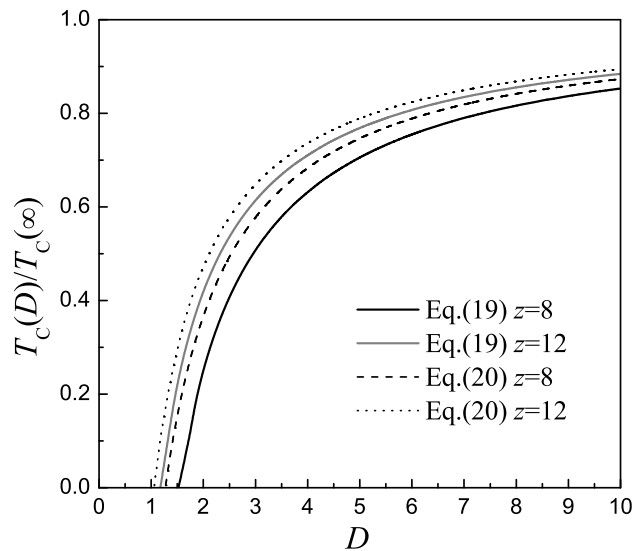}
 \caption{\label{fig03} Particle size dependence of the reduced critical temperature for pure ferromagnetic nanoparticles with $z$=8 and $z$=12, according to Eqs. (\ref{19}) and (\ref{20}).}
\end{figure}

As is observed, Curie temperature ($T_C$) decreases as the particle becomes smaller due basically to the decrease in the magnetic bonds density. Therefore, the energy cost to carry out the transition is lower, and thus the critical temperature is also smaller. As $D$ increases the critical temperature $T_C$ tends to its bulk value. Other models have been already proposed to understand the mechanisms lying on the effect of the breaking of exchange bonds upon the $T_C(D)$ function for nanoparticles. On this respect, the following expression has been proposed \cite{nikolaev}:
\begin{equation}\label{21}
\frac{T_C(D)}{T_C(\infty)}=1-\frac{3\Delta L}{2D},
\end{equation}
where $\Delta L$ is the thickness of surface layer, and it has been considered as a parameter to characterize the deficiency in the number of exchange bonds for atoms at the surface region of a nanoparticle.  However, this model can not reproduce successfully experimental data of magnetite nanoparticles of different size with a constant $\Delta L$, and hence, it has been suggested that $\Delta L$ should vary with the particle size, but such dependence has not been yet established \cite{nikolaev}.
Another model, based on the energy-equilibrium criterion between the spin-spin exchange interactions and the thermal vibration energy of atoms at the transition temperature and a size-dependent Debye temperature function, was developed in order to obtain both $T_C(D)$ and $T_N(D)$ of ferromagnetic ad antiferromagnetic nanocrystals. Such model yielded the following expression for nanoparticles \cite{lang}:
\begin{equation}\label{22}
\frac{T_C(D)}{T_C(\infty)}=exp\left[-(\alpha -1)/(D/D_0-1)\right],
\end{equation}
where $\alpha$ is a measure of the root-mean-square (rms) thermal average amplitude of surface atoms vibration relative to the core and $D_0$ denotes a critical size at which all atoms of the nanocrystal are located on its surface. Differently, our model contains just one adjustable parameter ($J$).  Concerning a comparison with experimental results, we want to stress that, in general, is rather difficult to obtain a diameter dependence of the Curie temperature due to several factors like shape inhomogeneities, size distribution, and, in some cases like Fe nanoparticles, surface oxidation \cite{gango}. Despite of that, our results are, qualitatively, in good agreement with some others reported for nanostructures \cite{lang,cui,du,sun}. In order to evaluate the reliability of our model, we have carried out a comparison with some experimental data available for Ni nanostructures \cite{cui,sun-searson}. To do this, we have employed Eq.(\ref{19}) with $z$=12 corresponding to a fcc lattice according to the crystalline structure of Ni, $T_C(\infty)$=631 K, a lattice parameter $a$=3.52 \AA, and we have also assumed that $J$=$J(\infty)$. Results are shown in Fig. 4 where we have also included the results from the models cited above. 
\begin{figure}[h]
 \centering
 \includegraphics[scale=0.85]{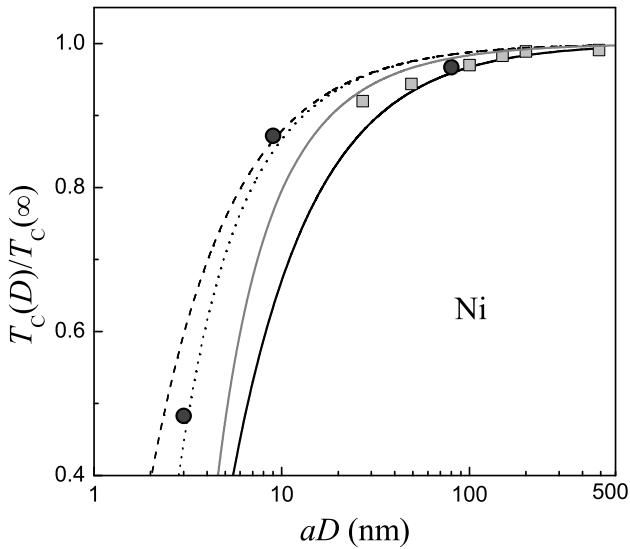}
 \caption{\label{fig04} Semi-log plots of the dependence of the Curie temperature on diameter for Ni nanostructures. Comparison between our model predictions (black solid line) without adjustable parameters according to Eq.(\ref{19}), available experimental results for Ni nanoparticles (circles, Refs.\cite{lang,cui}),  Ni nanorods (squares, Refs.\cite{lang,sun-searson}) and the theoretical models described in the text with $\Delta L$=0.8084 nm (dashed line, Eq.(\ref{21}), Refs. \cite{lang,nikolaev}) and $D_0$=1.4952 nm, $\alpha$=1.811 (dotted line, Eq.(\ref{22}), Ref.\cite{lang}). Gray line stands for our model using Eq.(\ref{23}) with $\alpha$=0.6.}
\end{figure}

As observed, the agreement is rather good despite of the simplicity of our model and without considering any adjustable free parameter. Discrepancies can be attributed to the fact that, firstly, we have considered a simple nearest neighbors Ising model. Second, experimental data correspond to Ni nanostructures that are not spherical at all, whereas our model has been developed for spherical nanoparticles. Third, for real nanostructures, the average lattice parameter is certainly different from bulk, mainly for very small particles of some few nanometers of diameter and, therefore, exchange integral should undergo changes. Such changes can also be induced by other facts like volume magnetostriction of Ni. Regarding a particle size dependence of an effective nearest neighbors exchange coupling, it is interesting to notice that better agreement with experimental data can be achieved by proposing a simple dependence of the form:
\begin{equation}\label{23}
J(D)=J(\infty)e^{\alpha/D},
\end{equation}
in Eq.(\ref{19}). This proposal is based on the experimental fact that the lattice parameter of metallic nanoparticles contracts with decreasing particle size in such a way that the lattice parameter contraction ($\Delta a/a$) is an inverse function of the diameter of nanoparticles \cite{frenkel,qi,jiang,chang,medasani}. Such a lattice contraction is attributed to reduction of surface bonds length as a response to surface stress. Contraction factors may vary with material and crystal orientation. It has been observed experimentally that the lattice parameter contracts by 2.4\% in 5 nm Ni particles \cite{stadnik}. On the other hand, fcc Ni is considered as a strong ferromagnet characterized by a less pronounced RKKY behavior, exponentially damped, and a faster decay of the exchange integral with the interatomic distance. More concretely, Ni remains ferromagnetic up to the 5$^{th}$ nearest neighbors, and within this range of distance the exchange integral is essentially a decreasing exponential function of interatomic spacing \cite{pajda,kaul}.

Concerning finite size scaling (FSS) properties, Fig. 5 shows a log-log plot of the reduced temperature $(T_C(\infty)-T_C(D))/T_C(\infty)$ versus particle diameter $D$, illustrating that the data obtained from Eq.(\ref{19}) follow the finite-size scaling relation \cite{fisher,landau-binder,binder,landau,privman}:
\begin{equation}\label{24}
\frac{T_C(\infty)-T_C(D)}{T_C(\infty)}=aD^{-1/\nu},
\end{equation}
from which our best estimate for the critical exponent associated to the correlation length is $\nu=1.0001\pm 0.0001$ for both bcc and fcc lattices. The observed exponent is slightly lower than the reported experimentally for Ni nanostructures ($\nu$=1.064)\cite{sun-searson}, and very similar to that of a two-dimensional (2D) Ising model ($\nu$=1) but much greater than the observed in the three-dimensional (3D) Ising model ($\nu$=0.6289)\cite{ferrenberg} and mean field theory ($\nu$=0.5), suggesting a 3D$\to$2D dimensionality crossover.
\begin{figure}[h]
 \centering
 \includegraphics[scale=0.81]{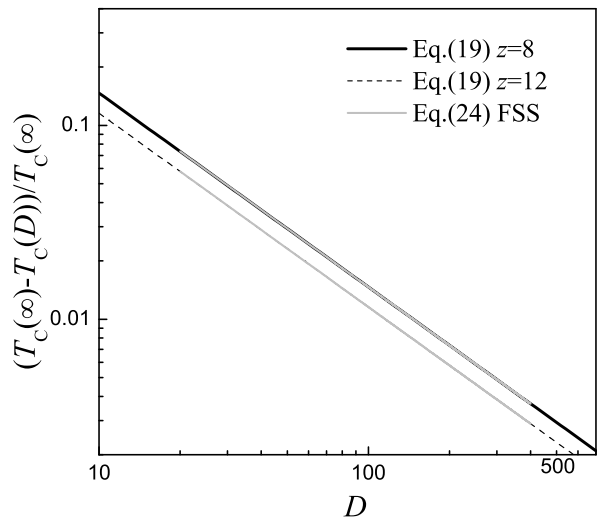}
 \caption{\label{fig05} Log-log plot of the size dependence of  the reduced temperature for  pure nanoparticles having core coordination numbers $z$=8 (bcc) (black solid line) and $z$=12 (fcc) (dashed line). Gray lines corresponds to the log-log fitting process using finite size scaling (FSS) theory (Eq.(\ref{24})).}
\end{figure}

Regarding thermal properties, Fig. 6 shows the temperature dependence of both the magnetization per site and the magnetic susceptibility for pure iron ($x$=0, $q$=0 and $p$=1) nanoparticles and for some selected diameters. Results for bulk iron are also included for comparison. A well-behaved thermal driven ferromagnetic to paramagnetic phase transition is observed as well as the shift of the critical temperature to low values as the system size decreases. The location of the maximum susceptibility coincides with that derived from Eq.(\ref{16}).
\begin{figure}[!hbtp]
 \centering
 \includegraphics[scale=0.84]{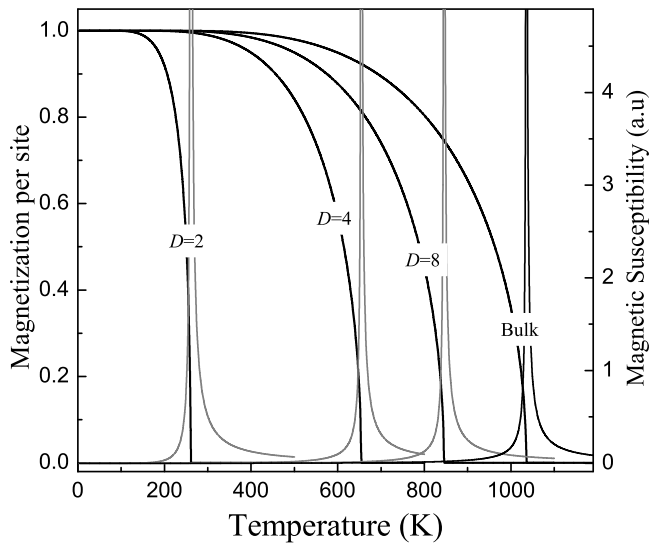}
 \caption{\label{fig06} Temperature dependence of both the magnetization per site and magnetic susceptibility for $\alpha$-Fe nanoparticles and bulk iron.}
\end{figure}

\subsection{FeMnAl nanoparticles}
In the  case of FeMnAl nanoparticles, we have chosen the stoichiometry Fe$_{0.5}$Mn$_{0.1}$Al$_{0.4}$, which, under bulk conditions, has been studied by using M\"ossbauer spectroscopy and magneometric techniques \cite{restre-perez} as well as from theory \cite{restrepo} and Monte Carlo simulation \cite{restre-greneche}. As is known, the bulk alloy has a $T_C$ close to room temperature ($\approx 300K$) in addition to the occurrence of a RSG behavior in the low temperature regime. The critical line $T_C(D)$ is given by Eq.(\ref{16}), and the corresponding log-log plot is shown in Fig. 7. Data have been fitted using finite size scaling theory (Eq.(\ref{24})). Our best estimate for the correlation length critical exponent is $\nu=0.926\pm 0.004$. This exponent is still quite similar to that of a 2D Ising model, but slightly different from our previous exponent for the pure case. This feature is consistent with the Harris's criterion \cite{harris1,harris2,landau1} for which a different set of critical exponents may be expected for diluted and random systems having a distribution of exchange integrals. In our case, dilution is provided by Al atoms and randomness is provided by the random distribution of the atomic elements in the alloy within the crystalline structure and over the entire volume of the nanoparticles. The exponent is also greater than the computed via Monte Carlo simulation ($\nu=0.79\pm 0.03$) of  Fe$_{0.5}$Mn$_{0.1}$Al$_{0.4}$ alloys under bulk conditions \cite{restre-greneche}. This fact agrees also with Fisher's theory for which critical behavior is modified when free-edge boundary conditions are considered giving rise to a new set of critical exponents different from those of the bulk \cite{fisher,landau2}. \\
In Fig. 8, we show the temperature dependence of magnetization per bond and magnetic susceptibility as obtained from Eqs.(\ref{14}) and (\ref{15}) respectively. As is observed, our model predicts a magnetization reduction in the low temperature regime, below around 70 K, in agreement with zero field cooling measurements for the bulk case \cite{restrepo}. According to our model, such a reduction, which has been attributed to a RSG phenomenology within the ferromagnetic matrix, is still observed for nanoparticles. Moreover, the onset of the reentrant phase is supported by the low temperature peaks of the magnetic susceptibility. Results reveal also the expected shift to lower temperature values of $T_C(D)$ as the particle size decreases, in agreement with the critical line obtained from Eq.(\ref{16}). 
\begin{figure}[!hbtp]
 \centering
 \includegraphics[scale=0.85]{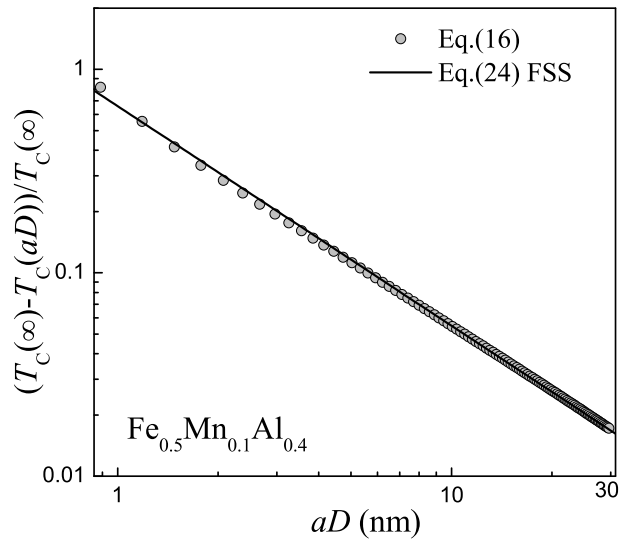}
 \caption{\label{fig07} Log-log plot of the size dependence of  the reduced temperature for Fe$_{0.5}$Mn$_{0.1}$Al$_{0.4}$ nanoparticles having core coordination numbers $z$=8 (bcc) (circles) using Eq.(\ref{16}). Solid line corresponds to the log-log fitting process for $z$=8 using finite size scaling (FSS) theory (Eq.(\ref{24})).}
\end{figure}

It is well established that Ising spin-glass transitions should follow the so-called Almeida-Thouless (AT) line \cite{almeida} from which is expected a field dependence of the peak temperature ($T_p$), obtained from the maximum of the magnetization, of the form:
\begin{equation}\label{25}
h\propto (1-T_p/T_f)^{3/2}.
\end{equation}
The extrapolation of the AT line at $h$=0 gives the freezing SG transition temperature $T_f$. Agreement of the data with the AT line is usually considered as evidence of the occurrence of a SG phase, although not concluding \cite{wenger}. Thus, in order to evaluate the properties of the RSG phase, we have solved Eq.(\ref{14}) for different low field values in Eq.(\ref{12}) from which $T_p$ was extracted with an uncertainty of $\pm$ 1 K. Results are summarized in Fig. 9 where we plot $h^{2/3}$ vs. $T_p$. Two remarkable features are observed in this figure. First, our data are in accordance with the AT line as one expects for a SG transition at least at low field values. The large plateau observed in the magnetization at around $T_p$ has been already observed to occur experimentally from SQUID and ac susceptibility measurements \cite{restre-perez}. Additionally, for large field values, a deviation from the A-T line was observed. These results are in agreement with those reported by Young et al. \cite{apyoung} and references therein where the difficulty of having an A-T line for short-range Ising spin glasses at large fields was evident. In fact, the existence of a SG ordering in a magnetic field is still an open question \cite{thomas}. Second, the freezing temperature is clearly size dependent, i.e. it diminishes as the particle size decreases. This fact implies that the SG region in the magnetic phase diagram becomes smaller for nanoparticles exhibiting SG behavior within their entire volume and not as a consequence of a merely surface effect as it has been proposed in nanoparticles exhibiting the so-called surface spin-glass-like behavior. These results suggest that SG behavior observed in bulk systems is reduced when finite size effects become important, which could be attributed to a reduction in the total number of frustrated spins as the particle size becomes smaller. Moreover, in the framework of the mean-field approximation, and taking into account that the low temperature transition occurs within the ordered Fe ferromagnetic matrix, the particle size dependence of the freezing temperature can be understood by writing $T_f= z_\text{eff}xJ_\text{Mn-Mn}+z_\text{eff}pJ_\text{Fe-Mn}$, which means that $T_f(D)$ should become proportional to $z_\text{eff}$, or at least as $1/D$ according to Eq.(\ref{11}). Inset in Fig. 9 reveals that such a trend is fulfilled.
\begin{figure}[h]
 \centering
 \includegraphics[scale=0.81]{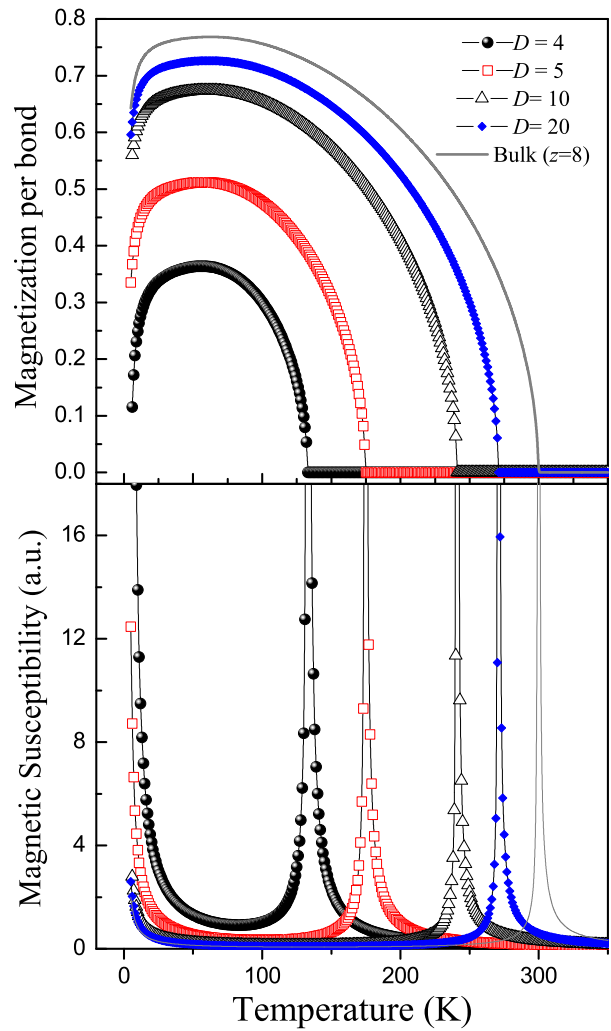}
 \caption{\label{fig08}Temperature dependence of magnetization per bond and magnetic susceptibility for Fe$_{0.5}$Mn$_{0.1}$Al$_{0.4}$ nanoparticles and for some selected particle sizes. Data corresponding to the bulk case are also included for comparison.}
\end{figure} \\
\begin{figure}[h]
 \centering
 \includegraphics[scale=0.85]{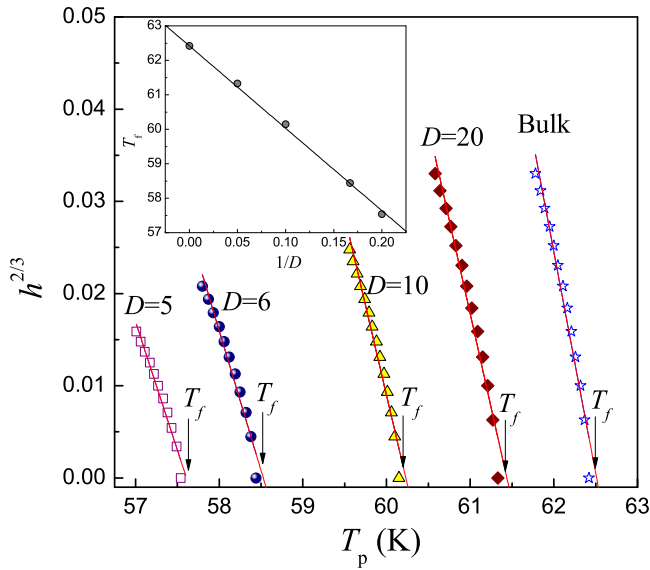}
 \caption{\label{fig09} Low field A-T line. Freezing temperature goes down as the particle size diminishes. The $T_f$ value for bulk at around 62.5 K is relatively close to that reported experimentally at around 78 K from ac susceptibility measurements for Fe$_{0.5}$Mn$_{0.1}$Al$_{0.4}$ bulk alloys \cite{restre-perez}. Inset shows that a $T_f$ versus $1/D$ linear dependence is followed.}
\end{figure} 

In order to interpret how such a low-temperature magnetization reduction takes place, we have performed a single-spin flip Metropolis Monte Carlo simulation \cite{metropolis,landau-binder} of Fe$_{0.5}$Mn$_{0.1}$Al$_{0.4}$ nanoparticles in the framework of a nearest-neighbor Ising model. We have used free boundary conditions, a maximum of 1$\times10^5$ Monte Carlo steps per spin (MCS) and discarded the first 6$\times10^4$ MCS for equilibration. Configurational averages over five different random atomic realizations were performed. We have also employed the same set of competitive parameters used in the variational approach and numerical values of $J_\text{Fe-Fe}$ \cite{restre-landau} reproducing the critical temperatures under bulk conditions. An example of the simulation results for a particle size $D$=10 is shown in Fig. 10.  In addition to the total magnetization per site, the corresponding Fe and Mn contributions  are shown separately. Both approaches, variational and simulational, predict a low temperature magnetization reduction. Monte Carlo results allow to conclude that such reduction arises from Mn moments for which an increase in the absolute value of the corresponding magnetization contribution was observed. These moments tend to align antiparallel with respect to the total magnetization direction ruled by the iron matrix according to the negative values of $J_\text{Fe-Mn}$ and $J_\text{Mn-Mn}$. Moreover, such moments are not compensated and some of them are frustrated.\\
\begin{figure}[h]
 \centering
 \includegraphics[scale=0.85]{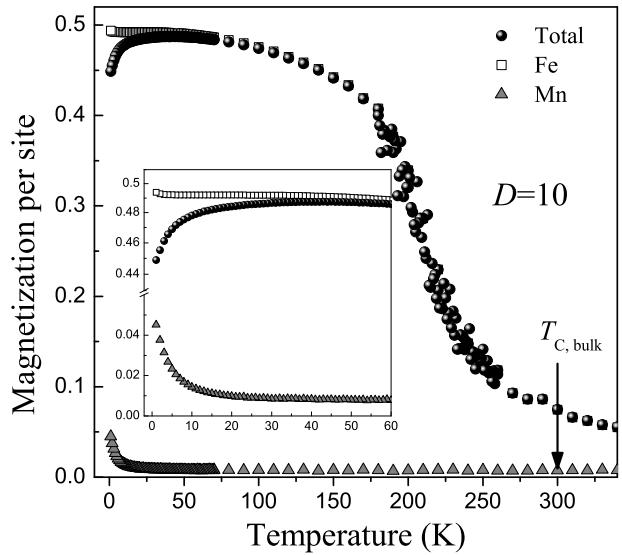}
 \caption{\label{fig10} Iron and manganese contributions to the total magnetization per site for Fe$_{0.5}$Mn$_{0.1}$Al$_{0.4}$ nanoparticles with diameter $D$=10. Inset shows a zoom of the low temperature behavior.}
\end{figure} 

One of the difficulties of the variational approach is that the magnetization obtained in the random diluted case is a magnetization per effective bond and not per atomic site. This fact gives rise to different values of the maximum magnetization (see Fig. 8) in contrast to the observed via Monte Carlo, where the maximum value of the overall magnetization is close to 0.5 accordingly with the Fe atomic concentration, which is practically the same independent of the particle size.
Finally, regarding the critical behavior along the $T_C$ line, we have determined the correlation length critical exponent from the maxima of the logarithmic derivative of the magnetization in the vicinity of $T_C$ by assuming the following ansatz \cite{ferrenberg}:
\begin{equation}\label{32}
\left(\frac{\partial\text{ln}m}{\partial T^{-1}}\right)_\text{max}=aD^{1/\nu}.
\end{equation}
The log-log plot of the size dependence of the maximum values of these derivatives is shown in Fig. 11 from which our best estimate for the exponent was $\nu$=0.71$\pm$0.04, very different from that found from the variational approach. Our value is however somewhat greater than the $\nu$=0.6289$\pm$0.0008 value obtained by Ferrenberg and Landau \cite{ferrenberg} for a 3D Ising model, where a high-resolution Monte Carlo study was carried out, and somewhat smaller than the $\nu$=0.79$\pm$0.03 value obtained for the same system under bulk conditions \cite{restre-greneche}. These comparisons suggest, contrary to the variational approach, that the 3D$\to$2D dimensionality crossover does not take place, at least for the range of $D$ values we have considered in the present study. The difference between the correlation length critical exponents, obtained via the variational approach and Monte Carlo, is attributed to the, so-called, pairs approximation in the variational approach where the trial partition function is computed by dividing the system only in blocks of one and two spins. This ends up in a reduction of the degree of correlation and consequently in a change of the exponent value. Certainly, a more precise estimative of the partition function, and consequently of the magnetization and $T_C$, can be achieved by considering the system as formed by higher order blocks (four, six, eight spins, etc) \cite{ferreira}. Even though such a calculation is tractable, it turns out heavy and very time consuming. This fact constitutes the main limitation of the variational model.\\  
On the other hand, differences respect to the $\nu$ exponent of the pure 3D Ising model can be attributed to the diluted character of our system in addition to the disorder involved in the distribution of exchange integrals, which is consistent with Harris's criterion \cite{harris1,harris2}, whereas the difference with respect to the $\nu$ exponent of the same system with periodic boundary conditions is consistent with Fisher's theory \cite{fisher}.
\begin{figure}[h]
 \centering
 \includegraphics[scale=0.85]{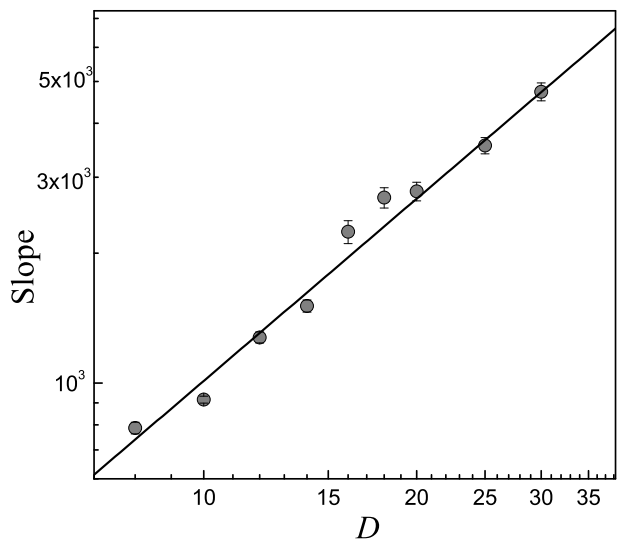}
 \caption{\label{fig11} Log-log plot of the particle diameter dependence of the maxima values of the logarithmic derivatives of the magnetization to determine the $\nu$ exponent.}
\end{figure} 

\section{Conclusions}
The critical behavior of ferromagnetic pure and random diluted nanoparticles with competing interactions has been addressed. In both cases we have employed the free energy variational principle based on the Bogoliubov inequality and an Ising model. In the case of random diluted nanoparticles, for which we have considered the Fe$_{0.5}$Mn$_{0.1}$Al$_{0.4}$ system as a case of study, we have used, additionally, standard Monte Carlo simulation. In order to validate the use of the variational approach in nanoparticles, which is carried out for the first time, the model was applied to account for the critical behavior of pure ferromagnetic nanoparticles on the basis of an average nearest neighbors coordination number obtained via numerical simulation. 
Our results allow to conclude that the variational model is successful in reproducing features of the particle size dependence of the Curie temperature for both pure and random diluted particles. Comparisons with other theoretical models and experimental results for Ni nanostructures were carried out in order to evaluate the reliability of the model. A better agreement with experimental data was obtained if a particle size dependence of the nearest neighbors exchange integral is considered consistent with previous works where lattice contraction of metallic nanoparticles has been observed.\\

For random diluted nanoparticles, low temperature magnetization reduction was consistent with the same type of RSG behavior observed in the bulk counterparts in accordance with the Almeida-Thouless line at low fields. Such a RSG behavior is attributed to the presence of competing interactions, randomness and the aluminum dilution effect. A linear dependence of the freezing temperature with the reciprocal of the particle diameter was also obtained indicating that the corresponding region in the magnetic phase diagram  becomes smaller as the particle size diminishes. Concerning critical behavior, data obtained by using the variational method were fitted with the relationship for critical temperature derived from finite size scaling theory and the best estimate for the correlation length critical exponent was $\nu$=0.926$\pm$0.004. Differently from this, a value $\nu$=0.71$\pm$0.04 was obtained via Monte Carlo. Differences are attributed to the so-called pairs-approximation in the variational approach. From both approaches, differences in the $\nu$ exponent of Fe$_{0.5}$Mn$_{0.1}$Al$_{0.4}$ nanoparticles with respect to that of the pure Ising model agree with Harris and Fisher arguments. Finally, we want to stress that, even though thermodynamical models can be indeed used in the study of nanostructures \cite{warda} and they can reproduce experimental features, special attention must be paid regarding critical behavior depending on the approximations of the model. 

\begin{acknowledgments}
This work was supported by the Sustainability projects of the Solid State and the
Microelectronic and Scientific Instrumentation Groups, and the IN565CE, IN576CE, IN578CE projects of the Antioquia University. J.R. thanks to the Spanish Education Ministry for the sabbatical year grant SB2009-0210 at the University of Barcelona. \`{O}.I. thanks financial support from Spanish MICINN through project MAT2009-0866 and Generalitat de Catalunya through project 2009SGR876. E.A.V. wants to thank Universidad San Buenaventura for financial support. We also acknowledge CESCA and CEPBA under coordination of C4 for computer facilities.
\end{acknowledgments}

\newpage 

\end{document}